\documentclass[12pt,epsf,epsfig]{article}
\usepackage{amsmath}

\begin{document}

\begin{center}
{\huge{Implications of the Spectrum of Dynamically Generated String Tension Theories}} \footnote{Based on an Essay written for the Gravity Research Foundation  2021 Awards for Essays on Gravitation, submission date : March 6th, 2021, awarded honorable mention} \\
\end{center}

\begin{center}
 E.I. Guendelman  \\
\end{center}

\begin{center}
\ Department of Physics, Ben-Gurion University of the Negev, Beer-Sheva, Israel \\
\end{center}

\begin{center}
\ Frankfurt Institute for Advanced Studies, Giersch Science Center, Campus Riedberg, Frankfurt am Main, Germany \\
\end{center}

\begin{center}
\ Bahamas Advanced Studies Institue and Conferences,  4A Ocean Heights, Hill View Circle, Stella Maris, Long Island, The Bahamas \\
\end{center}
E-mail:  guendel@bgu.ac.il,     

\abstract
The string tension does not have to be put in by hand, it can be dynamically generated, as in the case when we formulate string theory in the modified measure formalism, and other formulations as well. Then string tension appears, but as an additional dynamical degree of freedom . It can be seen however that this string tension is not universal, but rather each string generates its own string tension, which can have a different value for each string. We also define a new Tension scalar background field which change locally the value of the string tension along the world sheets of the strings. When there are many strings with different string tensions this Tension field can be determined from the requirement of world sheet conformal invariance and for two types of string tensions depending on the relative sign of the tensions we obtain non singular cosmologies and warp space scenarios and when the two string tensions are positive, we obtain scenarios where the Hagedorn temperature is avoided in the early universe or in regions of warped space time where the string tensions
become very big.

\section{Introduction}

String Theory is a candidate for the theory of all matter and interactions including gravity. But it has a dimension full parameter, the tension of the string, in its standard formulation.

The appearance of a dimension full string tension from the start appears somewhat unnatural. Previously however, in the framework of a Modified Measure Theory, a formalism originally used for gravity theories, see for example \cite{d,b}, the tension was derived as an additional degree of freedom \cite{a,c,supermod, cnish, T1, T2}. See also the treatment by Townsend and collaborators \cite{xx,xxx}.

This essay is organized as follows. In Section 2 we review the modified-measure theory in the string context. In Section 3 we discuss the fact that this tension generation could take place independently for each world sheet separately, which would mean that the string tension is not a fundamental coupling in nature and it could be different for different strings. In section 4 we discuss possible consequences for physics that are derived from these string theories with dynamical string tension generation. In particular we review the results obtained in when  we  define a new Tension scalar background field which change locally the value of the string tension along the world sheets of the strings. When there are many strings with different string tensions this Tension field can be determined from the requirement of world sheet conformal invariance of all the string world sheets. For two types of string tensions depending on the relative sign of the tensions we obtain non singular cosmologies and warp space scenarios, if the strings have opposite signs or, when the strings both are positive, we obtain scenarios where the Hagedorn temperature is avoided in regions of space time with very large string tensions (which means at those regions the Hagedorn temperature becomes infinity) in  cosmology and warp space scenarios. \\

\section{The Modified Measure Theory String Theory}

The standard world sheet string sigma-model action using a world sheet metric is:

\begin{equation}\label{eq:1}
S_{sigma-model} = -T\int d^2 \sigma \frac12 \sqrt{-\gamma} \gamma^{ab} \partial_a X^{\mu} \partial_b X^{\nu} g_{\mu \nu}.
\end{equation}

Here $\gamma^{ab}$ is the intrinsic Riemannian metric on the 2-dimensional string worldsheet and $\gamma = det(\gamma_{ab})$; $g_{\mu \nu}$ denotes the Riemannian metric on the embedding spacetime. $T$ is a string tension, a dimension full scale introduced into the theory by hand. \\

From the variations of the action with respect to $\gamma^{ab}$ and $X^{\mu}$ we get the following equations of motion:

\begin{equation} \label{eq:tab}
T_{ab} = (\partial_a X^{\mu} \partial_b X^{\nu} - \frac12 \gamma_{ab}\gamma^{cd}\partial_cX^{\mu}\partial_dX^{\nu}) g_{\mu\nu}=0,
\end{equation}

\begin{equation} \label{eq:3}
\frac{1}{\sqrt{-\gamma}}\partial_a(\sqrt{-\gamma} \gamma^{ab}\partial_b X^{\mu}) + \gamma^{ab} \partial_a X^{\nu} \partial_b X^{\lambda}\Gamma^{\mu}_{\nu\lambda}=0,
\end{equation}

where $\Gamma^{\mu}_{\nu\lambda}$ is the affine connection for the external metric. \\

There are no limitations on employing any other measure of integration different than $\sqrt{-\gamma}$. The only restriction is that it must be a density under arbitrary diffeomorphisms (reparametrizations) on the underlying spacetime manifold. The modified-measure theory is an example of such a theory. \\

In the framework of this theory two additional worldsheet scalar fields $\varphi^i (i=1,2)$ are introduced. A new measure density is

\begin{equation}
\Phi(\varphi) = \frac12 \epsilon_{ij}\epsilon^{ab} \partial_a \varphi^i \partial_b \varphi^j.
\end{equation}

Then the modified bosonic string action is (as formulated first in \cite{a} and latter discussed and generalized also in \cite{c})

\begin{equation} \label{eq:5}
S = -\int d^2 \sigma \Phi(\varphi)(\frac12 \gamma^{ab} \partial_a X^{\mu} \partial_b X^{\nu} g_{\mu\nu} - \frac{\epsilon^{ab}}{2\sqrt{-\gamma}}F_{ab}(A)),
\end{equation}

where $F_{ab}$ is the field-strength  of an auxiliary Abelian gauge field $A_a$: $F_{ab} = \partial_a A_b - \partial_b A_a$. \\

To check that the new action is consistent with the sigma-model one, let us derive the equations of motion of the action (\ref{eq:5}). \\

The variation with respect to $\varphi^i$ leads to the following equations of motion:

\begin{equation} \label{eq:6}
\epsilon^{ab} \partial_b \varphi^i \partial_a (\gamma^{cd} \partial_c X^{\mu} \partial_d X^{\nu} g_{\mu\nu} - \frac{\epsilon^{cd}}{\sqrt{-\gamma}}F_{cd}) = 0.
\end{equation}

It implies

\begin{equation} \label{eq:a}
\gamma^{cd} \partial_c X^{\mu} \partial_d X^{\nu} g_{\mu\nu} - \frac{\epsilon^{cd}}{\sqrt{-\gamma}}F_{cd} = M = const.
\end{equation}

The equations of motion with respect to $\gamma^{ab}$ are

\begin{equation} \label{eq:8}
T_{ab} = \partial_a X^{\mu} \partial_b X^{\nu} g_{\mu\nu} - \frac12 \gamma_{ab} \frac{\epsilon^{cd}}{\sqrt{-\gamma}}F_{cd}=0.
\end{equation}

We see that these equations are the same as in the sigma-model formulation (\ref{eq:tab}), (\ref{eq:3}). Namely, taking the trace of (\ref{eq:8}) we get that $M = 0$. By solving $\frac{\epsilon^{cd}}{\sqrt{-\gamma}}F_{cd}$ from (\ref{eq:a}) (with $M = 0$) we obtain (\ref{eq:tab}). \\

A most significant result is obtained by varying the action with respect to $A_a$:

\begin{equation}
\epsilon^{ab} \partial_b (\frac{\Phi(\varphi)}{\sqrt{-\gamma}}) = 0.
\end{equation}

Then by integrating and comparing it with the standard action it is seen that

\begin{equation}
\frac{\Phi(\varphi)}{\sqrt{-\gamma}} = T.
\end{equation}

That is how the string tension $T$ is derived as a world sheet constant of integration opposite to the standard equation (\ref{eq:1}) where the tension is put ad hoc.The variation with respect to $X^{\mu}$ leads to the second sigma-model-type equation (\ref{eq:3}). The idea of modifying the measure of integration proved itself effective and profitable. This can be generalized to incorporate super symmetry, see for example \cite{c}, \cite{cnish}, \cite{supermod} , \cite{T1}.
For other mechanisms for dynamical string tension generation from added string world sheet fields, see for example \cite{xx} and \cite{xxx}. However the fact that this string tension generation is a world sheet effect 
and not a universal uniform string tension generation effect for all strings has not been sufficiently emphasized before, this we do in our next section.

\section{Each String in its own world sheet determines its own string tension. Therefore the string tension is not universal for all strings}

Let us now observe indeed that it does not appear that the string tension derived above corresponds to ¨the¨ string tension of the theory. The derivation of the string tension in the previous section holds for a given string, another string could acquire a different string tension. 
Similar situation takes place in the dynamical string generation proposed by Townsend for example \cite{xx}, in that paper world sheet fields include an electromagnetic gauge potential. Its equations of motion are those of the Green-Schwarz superstring but with the string tension given by the circulation of the world sheet electric field around the string. So again ,in \cite{xx} also a string will determine a given tension, but another string may determine another tension. 
If the tension is a universal constant valid for all strings, that would require an explanation in the context of these dynamical tension string theories, for example some kind of interactions that tend to equalize string tensions, or that all  strings in the universe originated from the splittings of one primordial string or some other mechanism. 

In any case, if one believes in strings , in the light of the dynamical string tension mechanism being a process that takes place at each string independently, we must ask whether all strings have the same string tension. 

\section{The Tension Scalar and its Consequences}

 The string tension can be influenced by external fields for example \cite{Ansoldi}.
 Indeed, if to the action of the string  we add a coupling
to a world-sheet current $j ^{a}$,  i.e. a term
\begin{equation}
    S _{\mathrm{current}}
    =
    \int d ^{2} \sigma
        A _{a}
        j ^{a}
    ,
\label{eq:bracuract}
\end{equation}
 then the variation of the total action with respect to $A _{a }$
gives
\begin{equation}
    \epsilon ^{a b}
    \partial _{a }
    \left(
        \frac{\Phi}{\sqrt{- \gamma}}
    \right)
    =
    j ^{b}
    .
\label{eq:gauvarbracurmodtotact}
\end{equation}
Suppose that we have an external scalar field $\phi (x ^{\mu})$
defined in the bulk. From this field we can define the induced
conserved world-sheet current
\begin{equation}
    j ^{b}
    =
    e \partial _{\mu} \phi
    \frac{\partial X ^{\mu}}{\partial \sigma ^{a}}
    \epsilon ^{a b}
    \equiv
    e \partial _{a} \phi
    \epsilon ^{a b}
    ,
\label{eq:curfroscafie}
\end{equation}
where $e$ is some coupling constant. The interaction of this current with the world sheet gauge field  is also invariant under local gauge transformations in the world sheet of the gauge fields
 $A _{a} \rightarrow A _{a} + \partial_{a}\lambda $.

For this case,  (\ref{eq:gauvarbracurmodtotact}) can be integrated to obtain
\begin{equation}
  T =  \frac{\Phi}{\sqrt{- \gamma}}
    =
    e \phi + T _{i}
    ,
\label{eq:solgauvarbracurmodtotact2}
\end{equation}
or  equivalently
\begin{equation}
  \Phi
    =
   \sqrt{- \gamma}( e \phi + T _{i})
    ,
\label{eq:solgauvarbracurmodtotact}
\end{equation}
The constant of integration $T _{i}$ may vary from one string to the other. Notice tha the interaction is metric independent since the internal gauge field does not transform under the the conformal transformations. This interaction does not therefore spoil the world sheet conformal transformation invariance in the case the field $\phi$ does not transform under this transformation.  One may interpret 
(\ref{eq:solgauvarbracurmodtotact} ) as the result of integrating out classically (through integration of equations of motion) or quantum mechanically (by functional integration of the internal gauge field, respecting the boundary condition that characterizes the constant of integration  $T _{i}$ for a given string ). Then replacing 
$ \Phi
    =
   \sqrt{- \gamma}( e \phi + T _{i})$ back into the remaining terms in the action gives a correct effective action for each string. Each string is going to be quantized with each one having a different $ T _{i}$. The consequences of an independent quantization of  many strings with different $ T _{i}$
covering the same region of space time will be studied next.
 we can incorporate the result of the tension as a function of scalar field $\phi$, given as $e\phi+T_i$, for a string with the constant of integration $T_i$ by defining the action that produces the correct 
equations of motion for such string, adding also other background fields, the anti symmetric  two index field $A_{\mu \nu}$ that couples to $\epsilon^{ab}\partial_a X^{\mu} \partial_b X^{\nu}$
and the dilaton field $\varphi $ that couples to the topological density $\sqrt{-\gamma} R$
\begin{equation}\label{variablestringtensioneffectiveacton}
S_{i} = -\int d^2 \sigma (e\phi+T_i)\frac12 \sqrt{-\gamma} \gamma^{ab} \partial_a X^{\mu} \partial_b X^{\nu} g_{\mu \nu} + \int d^2 \sigma A_{\mu \nu}\epsilon^{ab}\partial_a X^{\mu} \partial_b X^{\nu}+\int d^2 \sigma \sqrt{-\gamma}\varphi R .
\end{equation}
Notice that if we had just one string, or if all strings will have the same constant of integration $T_i = T_0$.

In any case, it is not our purpose here to do a full generic analysis of all possible background metrics, antisymmetric two index tensor field and dilaton fields, instead, following \cite{cosmologyandwarped}
and \cite{noHagedorn} , 
we will take  cases where the dilaton field is a constant or zero, and the antisymmetric two index tensor field is pure gauge or zero, then the demand of conformal invariance for $D=26$ , see for example \cite{Polchinski}. becomes the demand that all the metrics, for two types of string tensions ($i=1,2$), 
\begin{equation}\label{tensiondependentmetrics}
g^i_{\mu \nu} =  (e\phi+T_i)g_{\mu \nu}
\end{equation}
will satisfy simultaneously the vacuum Einstein´s equations,\begin{equation}\label{Einstein1}
R_{\mu \nu} (g^1_{\alpha \beta}) = 0 
\end{equation}
and , at the same time,
\begin{equation}\label{Einstein1}
  R_{\mu \nu} (g^2_{\alpha \beta}) = 0
\end{equation}

These two simultaneous conditions above  impose a constraint on the tension field
 $\phi$, because the metrics $g^1_{\alpha \beta}$ and $g^2_{\alpha \beta}$ are conformally related, but Einstein´s equations are not conformally invariant, so the condition that Einstein´s equations hold  for both  $g^1_{\alpha \beta}$ and $g^2_{\alpha \beta}$
is highly non trivial.
Then for these situations, we have,
\begin{equation}\label{relationbetweentensions}
e\phi+T_1 = \Omega(e\phi+T_2)
\end{equation}
 which leads to a solution for $e\phi$
 
\begin{equation}\label{solutionforphi}
e\phi  = \frac{\Omega T_2 -T_1}{1 - \Omega} 
\end{equation}
which leads to the tensions of the different strings to be
\begin{equation}\label{stringtension1}
 e\phi+T_1 = \frac{\Omega(T_2 -T_1)}{1 - \Omega} 
\end{equation}
and
  \begin{equation}\label{stringtension2}
 e\phi+T_2 = \frac{(T_2 -T_1)}{1 - \Omega} 
\end{equation}

There are many solutions with $\Omega$, for example multiplying a Schwarschild solution by a constant, gives another Schwarschild solution of vacuum Einstein´s equations, the same is true with Kasner solutions.

Then there are solutions with two types of strings covering the same region of space time,  where $\Omega$ is not a constant, and in this case we differentiate between $\Omega$ positive and $\Omega$ negative.
 To first cosmological solutions it is  useful to consider flat space in the Milne representation, $D=4$ this reads,
 \begin{equation}\label{Milne4D}
 ds^2 = -dt^2 + t^{2}(d\chi^2 + sinh^2\chi d\Omega_2^2)
\end{equation}

where $ d\Omega_2^2 $ represent the contribution of the 2 angles to the metric when using spherical coordinates, that is, it represents the metric of a two dimensional sphere of unit radius. In $D$ dimensions we will have a similar expression but now we must introduce the metric of a  $D-2$ unit sphere $ d\Omega_{D-2}^2 $ so we end up with the following metric that we will take as the metric 2
 \begin{equation}\label{MilneD1}
 ds_2^2 = -dt^2 + t^{2}(d\chi^2 + sinh^2\chi d\Omega_{D-2}^2)
\end{equation}

For the metric $1$ we will take the metric that we would obtain from the coordinate $t \rightarrow 1/t $ (using Minkowskii coordinates 
$x^\mu $, this corresponds to the inversion transformation, for a review and generalizations see  \cite{Kastrup}.
 $x^\mu \rightarrow x^\mu/(x^\nu x_\nu ) $) and then we furthermore multiply by a constant $\sigma$, so

 \begin{equation}\label{MilneD2}
 ds_1^2 =\frac{\sigma}{t^4} (-dt^2 + t^{2}(d\chi^2 + sinh^2\chi d\Omega_{D-2}^2))
\end{equation}
Then the equations (\ref{relationbetweentensions}), (\ref{solutionforphi}), (\ref{stringtension1}), (\ref{stringtension2}),
with $ \Omega= \frac{\sigma}{t^4}$.
The strings 1 and 2 have both positive tensions if $\sigma $ is positive and if the sign of $T_2 -T_1$ is positive and the solution is not continued before the singularity which takes place when  $ \Omega= \frac{\sigma}{t^4} = 1$. At that point both string tensions apprach infinity and we have the possibility of avoiding the Hagedorn Temperature in the early universe.

If $\sigma$ is negative, we obtain a non singular cosmology with negative tensions dominating in the early universe and positive tensions dominating in the late universe. The metric $g_{\mu \nu}$ contains a bounce,
 \begin{equation}\label{universalmetric}
 ds^2 =g_{\mu \nu}dx^{\mu}dx^{\nu} =  (\frac{{1 - \Omega}}{T_2 -T_1})(-dt^2 + t^{2}(d\chi^2 + sinh^2\chi d\Omega_{D-2}^2))
\end{equation}
and considering that  $\Omega = \frac{\sigma}{t^4} = \frac{-K}{t^4}$ ,
where $K$ is positive. So the coefficient of the hyperbolic D -1 dimensional metric $d\chi^2 + sinh^2\chi d\Omega_{D-2}^2$ is 
$\frac{{t^{2} + \frac{-K}{t^2}}}{T_2 -T_1}$, showing a contraction, a bounce and a subsequent expansion. The initial and final spacetimes are flat and satisfy vacuum Einstein´s equations, but not the full space, with most appreciable deviations from   Einstein´s equations at the bouncing time, $t =t*= K^{1/4}$.

One can consider also warped spaces , as the ones discussed by Wesson,
these are solutions of hiher dimensions Einstein´s equations.  In  five dimensions for example the following warped solution is found,
\begin{equation}\label{Wesson1}
 ds^2 =l^2dt^2 -l^{2} cosh^{2}t (\frac{dr^2}{1-r^{2}} 
 + r^2 d\Omega_{2}^2)) - dl^{2}
\end{equation}
where $l$ is the fourth dimension,
so we see that as in the fourth dimension  $l$  such a solution is homogeneous of degree two, just as the Milne space time was homogeneous of degree two with respect to the time. Notice that maximally symmetric de Sitter space times sub spaces $l =$ constant appear for  instead of euclidean spheres that appear in the Milne Universe for $t =$ constant.

The list of space times of this type is quite large, for example, one cal find solutions of empty GR with Schwarzchild de Sitter subpaces for  $l =$ constant, as in
\begin{equation}\label{Wesson2}
 ds^2 =\frac{\Lambda l^{2}}{3} (dt^2 (1-\frac{2M}{r} -\frac{\Lambda r^2}{3}) - \frac{dr^2}{1-\frac{2M}{r} 
 -\frac{\Lambda r^2}{3}} 
 - r^2 d\Omega_{2}^2) - dl^{2}
\end{equation}
This of course can be extended to $D$ dimensions, where we choose one dimension $l$ to have a factor $l^2$ warp factor for the other dimensions , generically for  $D$ dimensions as in 
\begin{equation}\label{Wessongeneric}
 ds_2^2 =l^{2}\bar{g}_{\mu \nu}(x)dx^{\mu}dx^{\nu} - dl^{2}
\end{equation}
where $\bar{g}_{\mu \nu}(x)$ is a $D-1$ Schwarzschild de Sitter metric for example \cite{Wesson}. This we will take as our $2$ metric, 

In any case, working with this generic  metric of the form  (\ref{Wessongeneric}), but now in $D$ dimensions,  we can perform the inversion transformation $l \rightarrow \frac{1}{l} $, and multiplying also by a factor $ \sigma$ and obtain the conformally transformed metric $1$ that also satisfies the vacuum Einstein´s equations
\begin{equation}\label{Wessongenericinverted}
 ds_1^2 = \sigma l^{-2}\bar{g}_{\mu \nu}(x)dx^{\mu}dx^{\nu} - \sigma\frac{dl^{2}}{l^{4}} = \sigma l^{-4}ds_2^2
\end{equation}

From this point on , the equations the solutions for the tensions of the $1$ and $2$ strings have similar behavior to that in the cosmological case, just that $t  \rightarrow l$, so now $\Omega=  \sigma l^{-4}$, so that we now insert this  expression for $\Omega$ in (\ref{stringtension1}) and in  
(\ref{stringtension2}). 

Here, as in the cosmological case, we can distinguish two qualitatively different cases: one where the string tensions of the two types  of strings have different signs, that is $\sigma$ is negative, where there is no singularity for the metrics or the string tensions and where negative string tensions dominate at one limit of $l$ and positive string tensions dominate at one limit of $l$,

The other case is when $\sigma$ is positive, and we take the two types of string tensions positive,  here, there is a value  of $l$ where the string tensions approaches infinite, which means also that the Hagedorn temperature approaches infinity at this point, which means string can escape the Hagedorn phase transition by getting close to this value of  $l$, just as in the cosmological case also when $\sigma$ is positive, and we take the two types of string tensions positive
strings in the early universe are also able to escape the Hagedorn temperature.

One final comment on the methodology for finding the solutions: we indeed consider two conformally related space times  which are obtained from each other by a coordinate transformation. There is no local coordinate transformation that can bring the two spacetimes to the Minkowskii form simultaneously however. To see this one may consider the ratio $\frac{\sqrt{-g^1}}{\sqrt{-g^2}} = \Omega^{D/2}$, so $\Omega$ is a coordinate invariant and $\neq 1$. 

We notice finally that in the case that both the metric $1$ and the metric $2$ are two flat metrics, like in our studies of cosmological solutions, or for the warped spaces in the case the warped space  \ref{Wessongeneric} is defined with  $\bar{g}_{\mu \nu}(x)$ being de Sitter and not the more generic Schwarzschild de Sitter case, the solutions are most likely exact and not just to first order in the slope, since not only the Ricci tensor vanish but also the Riemann curvature and so will all higher curvature corrections to the beta function.

\textbf{Acknowledgments}
  I thank Oleg Andreev, David Andriot, Stefano Ansoldi,  David Benisty, Hitoshi Nishino, Emil Nissimov, Svetlana Pacheva, Subhash Rajpoot,  Euro Spallucci  and  Tatiana Vulfs for useful discussions. I also want to thank FQXi and the COST actions, CA18108 and CA16104 for support. \\


\begin{thebibliography}{9}
\bibitem{d}
E.I. Guendelman, A.B. Kaganovich, Phys.Rev.D55:5970-5980 (1997)
\bibitem{b}
E.I. Guendelman, Mod.Phys.Lett.A14, 1043-1052 (1999)
\bibitem{a}
E.I. Guendelman, Class.Quant.Grav. 17, 3673-3680 (2000)
\bibitem{c}
E.I. Guendelman, A.B. Kaganovich, E.Nissimov, S. Pacheva, Phys.Rev.D66:046003 (2002)
\bibitem{supermod}
E.I. Guendelman, Phys.Rev.D 63 (2001) 046006 • e-Print: hep-th/0006079 [hep-th]
\bibitem{cnish}
Hitoshi Nishino, Subhash Rajpoot, Phys.Lett.B 736 (2014) 350-355
e-Print: 1411.3805 [hep-th].
\bibitem{T1}
T.O. Vulfs, E.I. Guendelman, Annals Phys. 398 (2018) 138-145 • e-Print: 1709.01326 [hep-th]
\bibitem{T2}
T.O. Vulfs, E.I. Guendelman, Int.J.Mod.Phys.A 34 (2019) 31, 1950204 • e-Print: 1802.06431 [hep-th]
\bibitem{xx}
P.K. Townsend,  Phys.Lett.B 277 (1992) 285-288.
\bibitem{xxx}
E. Bergshoeff, L.A.J. London, P.K. Townsend,
Class.Quant.Grav. 9 (1992) 2545-2556, Class. Quantum Grav. 9 (1992) 2545-2556  • e-Print: hep-th/9206026 [hep-th]
\bibitem{Ansoldi}
S. Ansoldi, E. I. Guendelman, E. Spallucci, Mod.Phys.Lett.A 21 (2006) 2055-2065 • e-Print: hep-th/0510200 [hep-th]
\bibitem{cosmologyandwarped}   E.I. Guendelman, Cosmology and Warped Space Times in Dynamical String Tension Theories, e-Print: 2104.08875 [hep-th]
\bibitem{noHagedorn}   E.I. Guendelman, Escaping the Hagedorn Temperature in Cosmology and Warped Spaces with Dynamical String Tension, e-Print: 2105.02279 [hep-th] [hep-th]
\bibitem{Polchinski} Joseph Polchinski, String Theory, vol. 1 , Cambridge University Press (1998).
\bibitem{Wesson} Paul Wesson, ¨Space -Time-Matter. World Scientific, (1999); Paul S. Wesson, Hongya Liu, Int.J.Mod.Phys. D10 (2001) 905-912, • e-Print: gr-qc/0104045.

\end{thebibliography}
\end{document}